\documentclass[12pt]{article}

\usepackage[T2A]{fontenc}
\usepackage[utf8]{inputenc}
\usepackage[english]{babel}
\usepackage{graphicx}

\begin{document}

\title{\bfseries Separation of Potentials in the Two-Body Problem}

\author{\bfseries Andrey Vasilyev}

\date{Retired from State Optical Institute, Saint Petersburg, Russia
e-mail <andrey@wavemech.org>}

\maketitle

\begin{abstract}
\label{abstract}

In contrast to the well-known solution of the two-body problem through 
the use of the concept of reduced mass, a solution is proposed involving
separation of potentials. It is shown that each of the two point
bodies moves in its own stationary  potential well generated by the other body, and the magnitudes of these potentials are calculated. It is shown also that for each body separately the energy and the angular momentum 
laws are valid. The knowledge of the potentials in which 
the bodies are moving permits calculation of the
 trajectories of each body without resorting to the reduced mass.

\bigskip


\bigskip


\end{abstract}

\section{Introduction}

Consider the problem of motion of two point particles with masses 
$m_1$, $m_2$ in the absence of external forces 
(the two-body problem).
 Here and in what follows, all parameters belonging to the 
first particle will be labeled by index "1"\, and those belonging to the 
second particle, by "2". In order to avoid unnecessary complication of our 
consideration, we assume the center of mass of the system to be at rest. 
If the motion of the system is uniform, one can readily come to the 
corresponding conclusions by introducing a moving coordinate frame.

We shall look for the solution to the problem in the approximation used, 
for instance, in monograph  \cite{1}, i.e., in the nonrelativistic 
approximation.

There is a well known solution to this problem involving the reduced mass 
(cf., e.g., \cite{1}, \S 13 or \cite{2}, \S 12). In this case, analysis of 
the motion of two particles is replaced by consideration of the motion 
of one fictitious $\mu$-particle with a mass $\mu=\frac{m_1m_2}{m_1+m_1}$
(reduced mass), with this $\mu$-particle assumed to move in a centrally 
symmetric field with a center at rest placed at the center of mass of the
two particles. The magnitude of this field is governed by the force of 
interaction of the two particles with one another. On finding the law 
by which the $\mu$-particle moves, one will be able to readily reconstruct 
the motion of the real particles $m_1$ and $m_2$.

One can offer the other kind of solution. This  solution  assumes 
each particle moves in its own stationary potential well   and
for each body separately the energy and the angular momentum 
laws are valid. 
Said otherwise, in place of an 
analysis of the motion of one $\mu$-particle in a gravitational potential 
field associated
with particle interaction one may consider the motion of each real particle 
in the stationary gravitational potential field created by the other particle. 
This stationary field differs, however from the gravitational field produced
by the $m_1$ and $m_2$ masses. 
Our task is to fined this potential fields for every particle. 

\section{Finding the Gravitational Potentials for the Two-Body Problem}

Potential energy of interaction of two point masses at rest which are
on the distance $l=|\mathbf{l}|$  from one another is

\begin{equation}
\label{1}
\mathcal E_{Pot}(|\mathbf{l}|)=-G\frac{m_1m_2}{|\mathbf{l}|},
\end{equation} 
where $G$ is the gravitational constant, and $\mathbf{l}$ is the 
vector from particle "1"\ to particle "2"\ . This expression one
can write in another form 

\begin{equation}
\label{2}
\mathcal E_{Pot}(|\mathbf{l}|)=m_1\Phi_2^G(|\mathbf l|)=
m_2\Phi_1^G(|\mathbf l|).
\end{equation}
Here $\Phi_1^G(|\mathbf l|)$ and $\Phi_2^G(|\mathbf l|)$ are
gravitational potentials created by particles $m_1$ and $m_2$
respectively 

\begin{equation}
\label{3}
\Phi_1^G(|\mathbf l|)=-G\frac{m_1}{|\mathbf{l}|}, \qquad
\Phi_2^G(|\mathbf l|)=-G\frac{m_2}{|\mathbf{l}|}.
\end{equation}

Coordinate $|\mathbf l|$ is the distance between the particles. It is 
not convenient to use this coordinate when particles are moving. 
It is more  convenient to use  coordinate $\mathbf r$ which one can 
reckon from the center of mass. Taking it into account we write 
energies of the particles "1"\ and "2"\ , which are located at points 
$\mathbf{r}_1$ and $\mathbf{r}_2$ similar to formula (\ref{2})

\begin{equation}
\label{4}
\mathcal E_{1,Pot}(\mathbf{r}_1)=m_1\Phi_2(\mathbf{r}_1)= 
m_1K_2\Phi_2^G(\mathbf{r}_1),
\end{equation}

\begin{equation}
\label{5}
\mathcal E_{2,Pot}(\mathbf{r}_2)=m_2\Phi_1(\mathbf{r}_2)= 
m_2K_1\Phi_1^G(\mathbf{r}_2).
\end{equation}

Here $\Phi_1(\mathbf{r})$ and $\Phi_2(\mathbf{r})$ are the potentials 
which we must find,  $K_1$ and $K_2$ are the coefficients connected 
this potentials with the gravitational potentials generated by the
particles $m_1$ and $m_2$:

$$
\Phi_1(\mathbf{r})=K_1\left(-G\frac{m_1}{\mathbf{r}}\right), \qquad
\Phi_2(\mathbf{r})=K_2\left(-G\frac{m_2}{\mathbf{r}}\right).
$$

This formulation suggests that we are describing not the energy of 
interaction of particles with one another but rather the potential energy 
of each particle separately, which is referenced to a certain level. 
As is common practice, we take for the zero level the energy of interaction 
of a particle $m$  with other  particle at infinity. 

Let us turn now to the above potentials. We shall use for this purpose 
the relations derived for the well known solution involving the reduced mass.

We place the origin of coordinates at the center of mass of the system 
under consideration. In this case

\begin{equation}
\label{6}
m_1\mathbf{r}_1+m_2\mathbf{r}_2=0. 
\end{equation}
Introduce a vector defining the relative positions of the particles

\begin{equation}
\label{7}
\mathbf{l}=\mathbf{r}_2-\mathbf{r}_1.
\end{equation}
We have defined this vector by $\mathbf{l}$ rather than by $\mathbf{r}$
as this is done usually, in order not to confuse it with the running 
coordinate $\mathbf{r}$ which we reckon from the origin. Vector 
$\mathbf{l}$ is the same vector which enter in the (\ref{1}), (\ref{2})
and (\ref{3}) formulas.

The equalities (\ref{6}) and (\ref{7}) yield

\begin{equation}
\label{8}
\mathbf{r}_1=-\frac{m_2}{m_1+m_2}\mathbf{l},\qquad
\mathbf{r}_2=\frac{m_1}{m_1+m_2}\mathbf{l}.
\end{equation}

Differentiating Eqs. (\ref{6}) -- (\ref{8}) with respect to time, we come 
to similar relations for the point velocities

\begin{equation}
\label{9}
m_1\mathbf{v}_1+m_2\mathbf{v}_2=0, 
\end{equation}

\begin{equation}
\label{10}
\mathbf{v}=\mathbf{v}_2-\mathbf{v}_1,
\end{equation}

\begin{equation}
\label{11}
\mathbf{v}_1=-\frac{m_2}{m_1+m_2}\mathbf{v},\qquad
\mathbf{v}_2=\frac{m_1}{m_1+m_2}\mathbf{v}.
\end{equation}
Here $\mathbf{v}_1$ and $\mathbf{v}_2$ are the velocities of the particles
under consideration, and $\mathbf{v}$ is the relative particle velocity. 
Recall that it is the relative quantities $\mathbf{l}$ and $\mathbf{v}$
that are invoked in dealing with the problem with the use of reduced mass.
And it is the coordinate $\mathbf{l}$  that is employed to describe 
the trajectory of the $\mu$-particle.

We can use now Eq. (\ref{11}) to express the kinetic energy of each particle in terms of the relative velocity:

\begin{equation}
\label{12}
\mathcal E_{1,Kin}(\mathbf{r}_1)=\frac{m_1}{2}v_1^2(\mathbf{r}_1)=
\frac{m_1}{2}\frac{m_2^2}{(m_1+m_2)^2}v^2,
\end{equation}

\begin{equation}
\label{13}
\mathcal E_{2,Kin}(\mathbf{r}_2)=\frac{m_2}{2}v_2^2(\mathbf{r}_2)=
\frac{m_2}{2}\frac{m_1^2}{(m_1+m_2)^2}v^2.
\end{equation}

In Eqs. (\ref{12}) and (\ref{13}), the kinetic energy is expressed both 
through the velocities of individual particles and the relative velocity. 
An analysis of Eqs. (\ref{12}) and (\ref{13}) suggests that the expressions 
for the energy written in terms of the velocities of individual particles 
and through the relative velocity differ in characteristic factors:

\begin{equation}
\label{14}
\frac{m_1^2}{(m_1+m_2)^2}, \qquad \frac{m_2^2}{(m_1+m_2)^2}.
\end{equation}

We are going now to express the potential energy of each particle through 
its coordinates (see Eqs. (\ref{4}) and (\ref{5})), whereas up to now the potential energy of interaction of two particles was defined in terms of a relative parameter, namely, separation distance between the particles
 $|\mathbf{l}|$ (see Eq. (\ref{1})).

This suggests that a transition from description of the energy through 
separation distance between the particles, a relative parameter, to that in terms 
of the energy of each particle separately should bring about the appearance 
of factors of the kind of Eq. (\ref{14}).

Furthermore, by the virial theorem (see, e.g., \cite{1}, \S 10, or \cite{2},
\S 6), for the Coulomb potential well the kinetic energy is related to the 
potential energy through

$$
\bar\mathcal E_{Kin}=-\frac{1}{2}\bar\mathcal E_{Pot},
$$
where the line above denotes averaging over time. For circular motion, 
this relation is correct without averaging. Thus, the potential energy is 
proportional to the kinetic energy.

This all adds up to the following recasting of Eqs. (\ref{4})
and (\ref{5}):

\begin{equation}
\label{15}
\mathcal E_{1,Pot}(\mathbf{r}_1)=m_1\Phi_2(\mathbf{r}_1)=
m_1\frac{m_2^2}{(m_1+m_2)^2}\left(-G \frac{m_2}{|\mathbf{r}_1|}\right),
\end{equation}  

\begin{equation}
\label{16}
\mathcal E_{2,Pot}(\mathbf{r}_2)=m_2\Phi_1(\mathbf{r}_2)=
m_2\frac{m_1^2}{(m_1+m_2)^2}\left(-G\frac{m_1}{|\mathbf{r}_2|}\right).
\end{equation} 
We obtain

\begin{equation}
\label{17}
K_1=\frac{m_1^2}{(m_1+m_2)^2}, \qquad 
K_2=\frac{m_2^2}{(m_1+m_2)^2},
\end{equation}  
and potentials 

\begin{equation}
\label{18}
\Phi_1(\mathbf{r})=K_1\Phi_1^G(\mathbf{r})=
\frac{m_1^2}{(m_1+m_2)^2}\left(-G\frac{m_1}{|\mathbf{r}|}\right), 
\end{equation}  

\begin{equation}
\label{19}
\Phi_2(\mathbf{r})=K_2\Phi_2^G(\mathbf{r})=
\frac{m_2^2}{(m_1+m_2)^2}\left(-G\frac{m_2}{|\mathbf{r}|}\right).
\end{equation}  

Both potentials $\Phi_1(\mathbf{r})$ and $\Phi_2(\mathbf{r})$    form a stationary potential wells placed at the origin 
 in which the real particles $m_1$ and $m_2$ move. 
Each particle, however, moves in its “own”\ potential well, i.e., 
the particle $m_1$ moves in the potential well formed by the potential
 $\Phi_2(\mathbf{r})$,
and the particle $m_2$, in the potential well created by the potential
 $\Phi_1(\mathbf{r})$.

The  potentials $\Phi_1(\mathbf{r})$ and $\Phi_2(\mathbf{r})$
are naturally nothing more than conventional  potentials. 
They have been introduced by convention to describe the motion of the 
real particles  $m_1$ and $m_2$.

\section{Conservation Laws for Every Particle}

The time has come now to check whether the expressions for the potentials
 $\Phi_1(\mathbf{r})$ and $\Phi_2(\mathbf{r})$
 are correct. How can one verify the correctness of these relations?
First, the sum of the potential energies of both particles should be equal to the potential energy (\ref{1}) of interaction of the both  particles
with one another. 

Summing Eqs. (\ref{15}) and (\ref{16}) in conjunction with Eq. (\ref{8}), 
we see clearly that this equality is upheld for all $\mathbf{r}_1$
and $\mathbf{r}_2$ at any moment of time.

Because each particle moves in its fixed potential well, the laws of 
conservation of energy and momentum should be met for each particle separately.
Let us check it.

We use Eqs. (\ref{12}) and (\ref{13}) to find the ratio of the kinetic energies of the two particles

 \begin{equation}
\label{20}
\frac{\mathcal E_{1,Kin}(\mathbf{r}_1)}{\mathcal E_{2,Kin}(\mathbf{r}_2)}=
\frac{m_2}{m_1},
\end{equation} 
whence

\begin{equation}
\label{21}
\mathcal E_{1,Kin}(\mathbf{r}_1)=\frac{m_2}{m_1}
{\mathcal E_{2,Kin}(\mathbf{r}_2)}, \qquad
\mathcal E_{2,Kin}(\mathbf{r}_2)=\frac{m_1}{m_2}
{\mathcal E_{1,Kin}(\mathbf{r}_1)}.
\end{equation}
Next we take Eqs. (\ref{15}), (\ref{16}), and (\ref{8}) to find the ratio 
of the potential energies of the two particles

 \begin{equation}
\label{22}
\frac{\mathcal E_{1,Pot}(\mathbf{r}_1)}{\mathcal E_{2,Pot}(\mathbf{r}_2)}=
\frac{m_2}{m_1},
\end{equation}  
whence

\begin{equation}
\label{23}
\mathcal E_{1,Pot}(\mathbf{r}_1)=\frac{m_2}{m_1}
{\mathcal E_{2,Pot}(\mathbf{r}_2)}, \qquad
\mathcal E_{2,Pot}(\mathbf{r}_2)=\frac{m_1}{m_2}
{\mathcal E_{1,Pot}(\mathbf{r}_1)}.
\end{equation}  
Summing Eqs. (\ref{21}) and (\ref{23}) term by term, we come to the total
energies $\mathcal E_1(\mathbf{r}_1)$ and $\mathcal E_2(\mathbf{r}_2)$
of each particle

\begin{equation}
\label{24}
\mathcal E_1(\mathbf{r}_1)=
{\mathcal E_{1,Pot}(\mathbf{r}_1)}+
\mathcal E_{1,Kin}(\mathbf{r}_1)=\frac{m_2}{m_1}
({\mathcal E_{2,Pot}(\mathbf{r}_2)}+
\mathcal E_{2,Kin}(\mathbf{r}_2)),
\end{equation}
or, in a more concise form

\begin{equation}
\label{25}
\mathcal E_1(\mathbf{r}_1)=\frac{m_2}{m_1}
{\mathcal E_2(\mathbf{r}_2)}, \qquad
\mathcal E_2(\mathbf{r}_2)=\frac{m_1}{m_2}
{\mathcal E_1(\mathbf{r}_1)}.
\end{equation} 
Thus, we see that the ratio of the total energies of each particle obeys 
the equality similar to the relations (\ref{20}) and (\ref{22})

 \begin{equation}
\label{26}
\frac{\mathcal E_1(\mathbf{r}_1)}{\mathcal E_2(\mathbf{r}_2)}=
\frac{m_2}{m_1}.
\end{equation}  

The sum of the total energies of each particle should yield the total energy of the whole system. Using equalities (\ref{25}) we come to

 \begin{equation}
\label{27}
\mathcal E=\mathcal E_1(\mathbf{r}_1)+\mathcal E_2(\mathbf{r}_2)=
\mathcal E_1(\mathbf{r}_1)+\frac{m_1}{m_2}\mathcal E_1(\mathbf{r}_1)=
\mathcal E_1(\mathbf{r}_1)\frac{m_1+m_2}{m_2},
\end{equation}  

 \begin{equation}
\label{28}
\mathcal E=\mathcal E_1(\mathbf{r}_1)+\mathcal E_2(\mathbf{r}_2)=
\mathcal E_2(\mathbf{r}_2)+\frac{m_2}{m_1}\mathcal E_2(\mathbf{r}_2)=
\mathcal E_2(\mathbf{r}_2)\frac{m_1+m_2}{m_1}.
\end{equation}  

The total $\mathcal E$ is conserved as the energy of a closed system.
Hence, as follows from equalities (\ref{27}) and (\ref{28}), the total
energies of each of the particles are conserved, i.e., are coordinate 
independent

 \begin{equation}
\label{29}
\mathcal E_1(\mathbf{r}_1)=\mathcal E_1=const., \qquad
\mathcal E_2(\mathbf{r}_2)=\mathcal E_2=const.
\end{equation} 
Thus, each particle moves in its potential well with its own energy. 
 
Consider now the angular momentum of the system and of the particles separately.
We express the angular momentum of particles through the coordinates 
$\mathbf l$ and the relative velocity $\mathbf v$. Combining Eqs. (\ref{8}) 
and (\ref{11}), we write the angular momenta of the particles as follows

\begin{equation}
\label{30}
\mathbf M_1(\mathbf{r}_1)=m_1[\mathbf{r}_1\times\mathbf{v}_1]=
m_1\frac{m_2^2}{(m_1+m_2)^2}[\mathbf{l}\times\mathbf{v}],
\end{equation}  

\begin{equation}
\label{31}
\mathbf M_2(\mathbf{r}_2)=m_2[\mathbf{r}_2\times\mathbf{v}_2]=
m_2\frac{m_1^2}{(m_1+m_2)^2}[\mathbf{l}\times\mathbf{v}].
\end{equation}
As seen from Eqs. (\ref{30}) and (\ref{31}), the vectors 
$[\mathbf{r}_1\times\mathbf{v}_1]$,  $[\mathbf{r}_2\times\mathbf{v}_2]$ and 
$[\mathbf{l}\times\mathbf{v}]$ are oriented in the same direction.
From the ratio of the momenta

 \begin{equation}
\label{32}
\frac{\mathbf M_1(\mathbf{r}_1)}{\mathbf M_2(\mathbf{r}_2)}=
\frac{m_2}{m_1},
\end{equation}  
we immediately obtain

\begin{equation}
\label{33}
\mathbf M_1(\mathbf{r}_1)=\frac{m_2}{m_1}
{\mathbf M_2(\mathbf{r}_2)}, \qquad
\mathbf M_2(\mathbf{r}_2)=\frac{m_1}{m_2}
{\mathbf M_1(\mathbf{r}_1)}.
\end{equation}  

The sum of the momenta of the two particles should give the total angular 
momentum of the system. We sum the momenta to obtain

 \begin{equation}
\label{34}
\mathbf M=\mathbf M_1(\mathbf{r}_1)+\mathbf M_2(\mathbf{r}_2)=
\mathbf M_1(\mathbf{r}_1)+\frac{m_1}{m_2}\mathbf M_1(\mathbf{r}_1)=
\mathbf M_1(\mathbf{r}_1)\frac{m_1+m_2}{m_2},
\end{equation}  

 \begin{equation}
\label{35}
\mathbf M=\mathbf M_1(\mathbf{r}_1)+\mathbf M_2(\mathbf{r}_2)=
\mathbf M_2(\mathbf{r}_2)+\frac{m_2}{m_1}\mathbf M_2(\mathbf{r}_2)=
\mathbf M_2(\mathbf{r}_2)\frac{m_1+m_2}{m_1}.
\end{equation}  
The total angular momentum $\mathbf M$, as the momentum of a closed system, 
is conserved. But then, as seen from the equalities (\ref{34}) and 
(\ref{35}), the momenta of each of the particles are conserved, i.e., they are coordinate independent:

 \begin{equation}
\label{36}
\mathbf M_1(\mathbf{r}_1)=\mathbf M_1=const., \qquad
\mathbf M_2(\mathbf{r}_2)=\mathbf M_2=const.
\end{equation} 

Because for each particle both the total energy and the angular momentum 
are conserved, one may consider the motion of each particle in its potential well separately, and calculate the energies and angular momenta also separately for each particle. Also, there is no need to use the reduced mass; indeed, one may use instead real masses  of each particle. 
Relation of energies of the particles must obey Eqs.
(\ref{26}),  (\ref{27}) and (\ref{28}), and relation of angular momenta --
Eqs. (\ref{32}),  (\ref{34}) and (\ref{35}).

If, however, one has to know the positions of particles with respect to 
one another, one will have to consult Eqs. (\ref{6}) and (\ref{9}).

We note in conclusion that this approach permits one to separate completely
not only the kinetic but the potential energies as well. Indeed, instead of
expressing the potential energy through the relative coordinate 
$\mathbf{l}$ 
(see Eq. (\ref{1})), it can be defined in terms of the coordinates of each 
particle separately
$$
\mathcal E=\mathcal E_1+\mathcal E_2=
\mathcal E_{1,Kin}(\mathbf{v}_1)+\mathcal E_{1,Pot}(\mathbf{r}_1)+
\mathcal E_{2,Kin}(\mathbf{v}_2)+\mathcal E_{2,Pot}(\mathbf{r}_2).
$$

\section{Trajectories of Particle Motion}

Because in each potential well particles move saving its energies and
angular momenta
one can study the behavior of particles with the use of the well known 
relations describing the motion of one particle in a potential well 
(see, e.g., \cite{1}, \S 15). As follows from these relations, if the 
potential energy in a central field is inversely proportional to $r$
($r$ is the radial coordinate), i.e., it can be written as 
(showing explicitly the sign)

\begin{equation}
\label{37}
\mathcal E_{Pot}(r)=-\frac{\alpha}{r}, 
\end{equation} 
where $\alpha$ is a positive constant, then the particle trajectory 
in polar coordinates $r$, $\varphi$ has the shape of an ellipse with the 
focus at the origin

\begin{equation}
\label{38}
r=\frac{p}{1+e\cos(\varphi-c)}.
\end{equation} 
Here  $c$ is the constant of integration, and $p$ and $e$ are the 
orbital parameter and the eccentricity

\begin{equation}
\label{39}
p=\frac{M^2}{m\alpha}, 
\end{equation}

\begin{equation}
\label{40}
 e=\sqrt{1+\frac{2{\mathcal E}M^2}{m\alpha^2}}.
\end{equation}
Here $\mathcal E$ is the total energy of the particle, and  $M^2$ is the
squared angular momentum. If it is necessary to find the trajectory of  the particle 
one must know quantities $\mathcal E$ and $M^2$.

 It is known that the same relations 
can be used if one analyzes the relative motion of two particles. In this
case, one has to put for mass $m$ in these relations the reduced mass 
$\mu=\frac{m_1m_2}{m_1+m_2}$. A fictitious particle with mass $\mu$
describes the motion of a system of two particles. We are going in what
follows to label all quantities related to the fictitious particle $\mu$
by index $\mu$.

Let us see how the trajectories of each particle are related to that of 
the fictitious particle $\mu$ describing the behavior of a system of two 
particles. And also, whether it is possible to trace the possible 
trajectories of each particle without resorting to the intermediate $\mu$
particle.

If the axis from which the angle $\varphi$ is reckoned passes through 
the point of the ellipse closest to the focus (the perihelion), the constant 
of integration $c$ in Eq. (\ref{38}) is zero, and if this point is farthest 
from the focus of the ellipse, $c=\pm\pi$ (with either of the signs).
The perihelia of the orbits of particles “1”\ and “2”\ look, however, in 
opposite directions. This is seen clearly already from equality (\ref{6});
indeed, for this equality to hold, vectors $\mathbf r_1$ and $\mathbf r_2$
must reach the maximum and the minimum simultaneously. The orbits of the
particles are shown in Fig.~1. 

\begin{figure}
\includegraphics[scale=0.8]{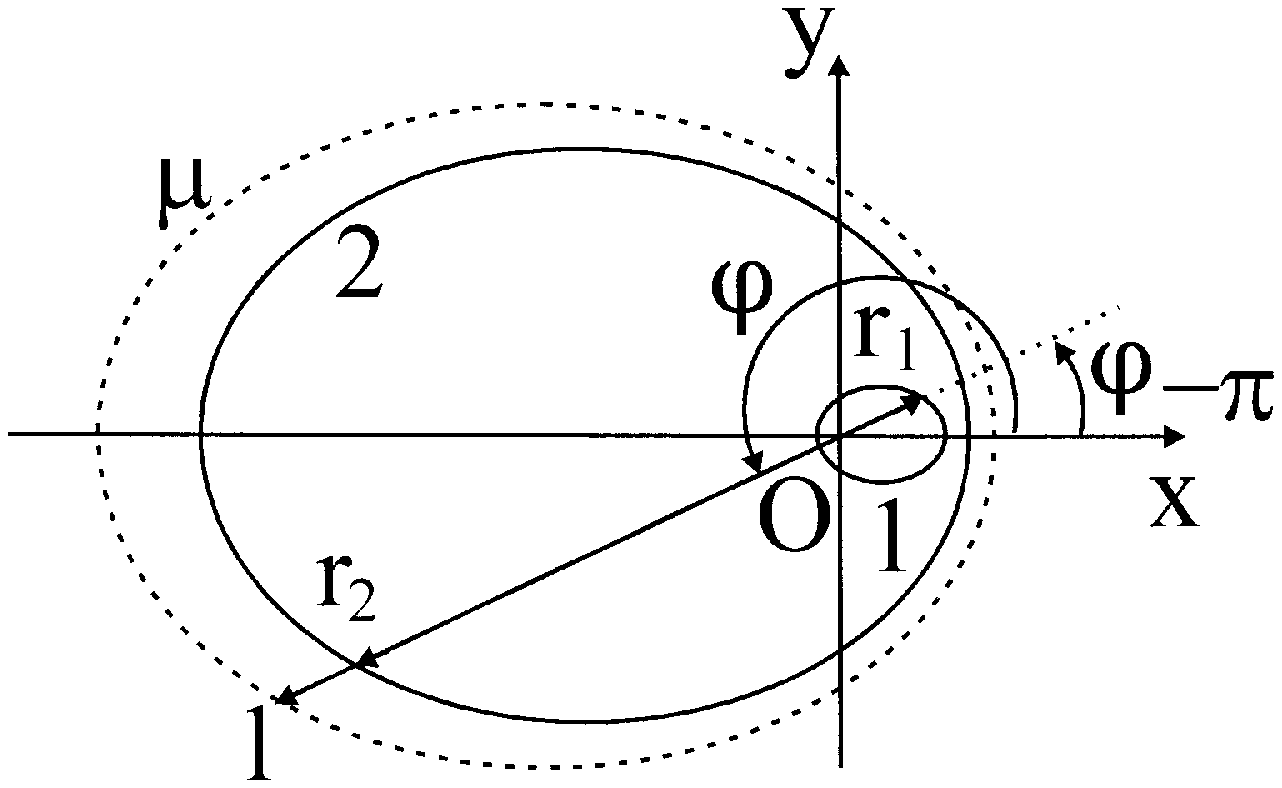}
\caption{}
\end{figure}

In Fig.~1, the center of mass of the system 
is at point $O$ (the origin). Also shown are the orbits of particles “1”\
and “2”\ , vector $\mathbf l=\mathbf r_2-\mathbf r_1$ and the $\mu$ orbit 
(dashed) along which the fictitious particle with mass $\mu$ moves. 
As seen from Fig.~1, the constant $c$ for the ellipse “1”\ should be equal
to  $c_1=\pm\pi$, and for the ellipse “2”\ ,  $c_2=0$.

We are going to demonstrate that the above approach offers a possibility 
of calculating possible trajectories of each particle without resorting 
to the reduced mass.

Apply now Eqs. (\ref{37})--(\ref{40}) to a study of the motion of our 
particles “1”\ and “2”\ . As seen from Eqs. (\ref{15}) and (\ref{16}), 
the parameter $\alpha$ for our particles can be written

\begin{equation}
\label{41}
\alpha_1=\frac{Gm_1m_2m_2^2}{(m_1+m_2)^2}, \qquad 
\alpha_2=\frac{Gm_1m_2m_1^2}{(m_1+m_2)^2},
\end{equation}
whence

\begin{equation}
\label{42}
\frac{\alpha_1}{\alpha_2}=\frac{m_2^2}{m_1^2}, \qquad
\alpha_1=\frac{m_2^2}{m_1^2}\alpha_2, \qquad
\alpha_2=\frac{m_1^2}{m_2^2}\alpha_1.
\end{equation}
Using Eqs. (\ref{26}), (\ref{32}), and (\ref{42}), one can readily verify
that $e_1=e_2$, i.e., the eccentricities for both particles are equal. 
This implies that the shape of the trajectory (shape of the ellipse) is 
the same for the two particles. As for the $p$ parameters, i.e., the
dimensions of the ellipses, these parameters obey the relation

\begin{equation}
\label{43}
\frac{p_1}{p_2}=\frac{m_2}{m_1},
\end{equation}
whence
\begin{equation}
\label{44}
p_1=\frac{m_2}{m_1}p_2, \qquad p_2=\frac{m_1}{m_2}p_1
\end{equation}

Accordingly, all the parameters related to the ellipse dimensions 
(the major and minor semiaxes and others) obey the same relation.

Express vector $\mathbf l$ from Eq. (\ref{7}) through the ellipse parameters 
(the coordinate $\mathbf l$ describes the trajectory of the $\mu$ particle). 
We can use the general relation (\ref{38}) to express $\mathbf r_1$ and 
$\mathbf r_2$ through the parameters of the ellipses. Because vectors 
$\mathbf r_1$ and $\mathbf r_2$ are antiparallel, in polar coordinates the 
vectors will differ by $\pi$. For the coordinate $r$ of the trajectory of 
the $\mu$ particle we have  $r=|\mathbf l|=
|\mathbf r_2-\mathbf r_1|=|\mathbf r_2|+|\mathbf r_1|$. The angular coordinate
of vector $\mathbf r_2$  is  $\varphi_2=\varphi$, because it is for the
“2”\ ellipse that the constant $c_2=0$. For the vector $\mathbf r_1$, 
the angular coordinate $\varphi_1=\varphi-\pi$. Apart from this, as already 
mentioned, for the ellipse “1”\ the constant $c_1=\pm\pi$. Using Eqs. 
(\ref{44}) and recalling that the directions of the vectors differ by $\pi$,
we come to

$$
r=r_\mu=\frac{p_2}{1+e\cos\varphi}+
\frac{p_1}{1+e\cos((\varphi-\pi)\pm\pi)}=\frac{m_1}{m_2}
\frac{p_1}{(1+e\cos\varphi)}+\frac{p_1}{1+e\cos\varphi},
$$
or

\begin{equation}
\label{45}
r_\mu=\frac{m_1+m_2}{m_2}\frac{p_1}{(1+e\cos\varphi)}, \qquad
r_\mu=\frac{m_1+m_2}{m_1}\frac{p_2}{(1+e\cos\varphi)}.
\end{equation}

Thus, the trajectory of the $\mu$ particle is also an ellipse, with the
same eccentricity as those of particles “1”\ and “2”\ , and with an orbit 
parameter (denote it by $p_\mu$)

\begin{equation}
\label{46}
p_\mu=\frac{m_1+m_2}{m_2}p_1, \qquad p_\mu=\frac{m_1+m_2}{m_1}p_2.
\end{equation}

Express the parameters of the ellipse $\mu$ in terms of those of ellipses
“1”\ and “2”\ . The angular momentum of the system should be the sum of
momenta of the particles and equal to the momentum of the $\mu$-particle,
because it is the latter that describes the behavior of the system:

$$
\mathbf M=\mathbf M_\mu=\mathbf M_1+\mathbf M_2. 
$$
We have already calculated the angular momentum of the system 
(Eqs. (\ref{34}) and (\ref{35})). Squaring $\mathbf M_\mu$ and using 
equality (\ref{33}), we come to

\begin{equation}
\label{47}
\mathbf M_\mu^2=\mathbf M_1^2+\mathbf M_2^2+2\mathbf M_1\mathbf M_2=
\mathbf M_1^2+\mathbf M_2^2+2\frac{m_1}{m_2}\mathbf M_1^2=
M_1^2+\frac{m_1^2}{m_2^2}M_1^2+2\frac{m_1}{m_2}M_1^2.
\end{equation}
Recasting expression (\ref{39}) to the form $M_1^2=p_1m_1\alpha_1$, we 
substitute in expression (\ref{47}) $p_1$ from relation (\ref{46}), 
and $\alpha_1$ from relation (\ref{41}), and take into account that
for system of two particles $\alpha=Gm_1m_2$. The end result is

\begin{equation}
\label{48}
\mathbf M_\mu^2=p_\mu\frac{m_1m_2}{m_1+m_2}Gm_1m_2=
p_\mu\mu\alpha.
\end{equation}

Thus we have once again come to the well known result, namely, if we express
the parameters of the ellipse along which the $\mu$-point moves through 
the parameters of motion of real particles, then for the mass we should 
take $\mu=\frac{m_1m_2}{m_1+m_2}$, i.e., the reduced mass.

Summing up, we have shown that in the “two-body problem”\ one can operate
without using the concept of the reduced mass. In this case, however, 
one should use not the energy of  particle interaction (\ref{37}) but 
rather the energy of each particle, which is reckoned from the zero level:

\begin{equation}
\label{49}
\mathcal E_{1,Pot}(r)=-\frac{Gm_1m_2m_2^2}{(m_1+m_2)^2}\frac{1}{r}, \qquad 
\mathcal E_{2,Pot}(r)=-\frac{Gm_1m_2m_1^2}{(m_1+m_2)^2}\frac{1}{r}.
\end{equation}
For the zero level one takes here the energy of interaction between 
particles infinitely far from one another.

Energies $\mathcal E_{1,Pot}(r_1)$ and $\mathcal E_{2,Pot}(r_2)$ are not
independent. We use formula (\ref{6}) of the relation of particles 
coordinates $\mathbf r_1$ and $\mathbf r_2$ to find relation
of potential energies of both particles

 \begin{equation}
\label{50}
\frac{\mathcal E_{1,Pot}({r_1})}{\mathcal E_{2,Pot}({r_2})}=
\frac{m_2}{m_1}.
\end{equation}  
Naturally, this expression  coincide with the equation  (\ref{22}).
If it is necessary to know not the energy only but the shape of the
trajectories one must take into account relation  (\ref{32}).

It might seem at first glance that this complicates solution of the problem.
This approach permits one, however, to separate the potential energies and
consider separately the motion of each particle in its potential well.

We note in conclusion that using the standard approach with the reduced mass
requires a two-step procedure, in which one first finds the solution for the 
reduced mass, and after that, derive from this solution the parameters of
motion of the particles of interest. Application of the above technique 
permits one to calculate {\itshape{\bfseries directly}} the possible 
trajectories of {\itshape{\bfseries each}} particle.

Such approach one can use in the case when particles have electric 
charges $q_1$ and $q_2$ (in the case under consideration charges
must have opposite signs). Such approach can be used only if velocities 
of the charges $v\ll c$ ($c$ is the light velocity). In this case 
we can neglect by the energy of the magnetic field which appears in
moving charged particles in comparison with the energy of the
electric field. Said otherwise, in this case we neglect by   the
vector potential $\mathbf A(\mathbf r)$ and claim that particles
move in scalar potential  $\Phi(\mathbf r)$ only.

As a rule in this case one can neglect by the gravitational interaction
in comparison with the electromagnetic one. Then instead the gravitational 
potential

$$
\Phi^G(\mathbf r)=\left(-G\frac{m}{|\mathbf r|}\right)
$$
it is necessary to put  electric field potential in all formulas (we use the 
Gaussian absolute system of units)

$$
\Phi(\mathbf r)=\frac{q}{|\mathbf r|}.
$$
Then quantity $\alpha$ from the Eq. (\ref{37}) is $\alpha=|q_1||q_2|$,
quantities $\alpha_1$ and $\alpha_2$ from the Eq.~(\ref{41}) are:

\begin{equation}
\label{51}
\alpha_1=\frac{|q_1||q_2|m_2^2}{(m_1+m_2)^2},   \qquad
\alpha_2=\frac{|q_1||q_2|m_1^2}{(m_1+m_2)^2}.
\end{equation}  
Energies of the particles can be written as (see Eq.~(\ref{49}))

\begin{equation}
\label{52}
\mathcal E_{1,Pot}(r_1)=-\frac{|q_1||q_2|m_2^2}{(m_1+m_2)^2}\frac{1}{r_1},   \qquad
\mathcal E_{2,Pot}(r_2)=-\frac{|q_1||q_2|m_1^2}{(m_1+m_2)^2}\frac{1}{r_2}.
\end{equation}  

We use Eq.~(\ref{6}) to find ratio of the potential energies of the 
two particles in this case

\begin{equation}
\label{53}
\frac{\mathcal E_{1,Pot}(r_1)}
{\mathcal E_{2,Pot}(r_2)}=\frac{m_2}{m_1}.
\end{equation}   
Hence in this case ratio of the potential energies of the 
two revolving particles are inversely proportional to their mass also. 

As was shown above coordinate $r$ of the trajectory  of the
$\mu$-particle is $r=r_\mu=|\mathbf l|$. We use Eq.~(\ref{8}) to sum the 
potential energies of every particle:

\begin{equation}
\label{54}
\mathcal E_{1,Pot}(r_1)+\mathcal E_{2,Pot}(r_2)=-\frac{|q_1||q_2|}{l}=
-\frac{|q_1||q_2|}{r_\mu}.
\end{equation}   
We receive Eq.~(\ref{37}) that is the energy of interaction of the two particles.

\end{document}